\documentclass[12pt,preprint]{aastex}
\usepackage{url}
\usepackage{natbib}
\usepackage{aas_macros}
\shorttitle{Mechanism of local dynamo action on the Sun}

\title{Mechanism of local dynamo action on the Sun}

\author{I.~N. Kitiashvili$^{1,2,3}$, A.~G. Kosovichev$^{1,4,5}$, N.~N. Mansour$^{2,6}$, A.~A. Wray$^{2,6}$}
\affil{$^1$Hansen Experimental Physics Laboratory, Stanford University, Stanford, CA 94305, USA}
\altaffiltext{1}{e-mail: irinasun@stanford.edu}
\affil{$^2$Center for Turbulence Research, Stanford University, Stanford, CA 94305, USA}
\affil{$^3$Kazan Federal University, Kazan, 420008, Russia}
\affil{$^4$Big Bear Solar Observatory, New Jersey Institute of Technology, Newark, NJ 07102, USA}
\affil{$^5$Crimean Astrophysical Observatory, Kiev National University, Nauchny, Crimea, 98409 Ukraine}
\affil{$^6$NASA Ames Research Center, Moffett Field, Mountain View, CA 94035, USA}

\begin{document}
\begin{abstract}
In the quiet Sun, magnetic fields are usually observed as small-scale magnetic elements, `salt and pepper', covering the entire solar surface. By using 3D radiative MHD numerical simulations we demonstrate that these fields are a result of local dynamo action in the top layers of the convection zone, where extremely weak `seed' magnetic fields can locally grow above the mean equipartition field (e.g., from a $10^{-6}$~G `seed' field to more than 1000~G magnetic structures). We find that the local dynamo action takes place only in a shallow, about 500~km deep, subsurface layer, from which the generated field is transported into deeper layers by convection downdrafts. We demonstrate that the observed dominance of vertical magnetic fields at the photosphere and the horizontal fields above the photosphere can be explained by multi-scale magnetic loops produced by the dynamo.
\end{abstract}

\section{Introduction}
Magnetic field generation is a key problem in understanding solar variability across wide range of scales. Modern high-resolution observations of the global magnetic field distribution, such as from HMI/SDO \citep{Scherrer2012},  and magnetic fields in selected areas by NST/BBSO \citep{Goode2010}, SOT/Hinode \citep{Tsuneta2008} and IMAX/SUNRISE \citep{Solanki2010} demonstrate the  complicated dynamics of magnetic fields and a tendency to self-organization. Traditionally, the solar dynamo problem is divided according to two scales: the global dynamo, operating on the scale of the 22-year solar cycle and controlling the global field \citep[e.g.][]{Choudhuri1995,Dikpati1999,Brandenburg2012}, and the local dynamo \citep[e.g.][]{Cattaneo1999,Vogler2007}, which operates on the scale of granulation and super-granulation and is believed to be responsible for the Sun's `magnetic carpet' \citep{Schrijver1998,Schrijver2002}. However, energetic and dynamic connections between the two dynamos are not clear. Recently, it became possible to resolve flows and magnetic fields associated with magnetic flux emergence on very small subgranular scales \citep[e.g.][]{Centeno2007,OrozcoSuarez2008,MartinezGonzalez2009,Thornton2011}, which probably reflects the working of the local dynamo in subsurface layers. Resolving magnetic fluxes though observations on smaller and smaller scales has renewed interest in small-scale magnetic field generation and raised a question about the existence of unresolved (or `hidden') magnetic flux in the quiet Sun \citep[e.g.,][]{Stenflo1982,Stenflo2012,TrujilloBueno2004,Shchukina2011}.

Numerical simulation is an efficient way to investigate properties of magnetic field generation on small scales. In particular, direct numerical simulations of simplified convective flows demonstrate the existence of the local dynamo and provide its basic characteristics, for instance, the effect of the magnetic Prandtl number on dynamo properties in the case of general turbulent flows \citep[e.g.,][]{Meneguzzi1989,Schekochihin2004,Schekochihin2005} and also for solar-type convection \citep[e.g.,][]{Cattaneo1999,Brandenburg2012}. In addition, recent `realistic'-type radiative MHD simulations have reproduced solar conditions with a high degree of realism and demonstrated that the magnetic field can be quickly amplified by local dynamos from a $10^{-2}$~G seed field to $\sim 25$~G magnetic elements \citep{Vogler2007,PietarilaGraham2010}, and a 1~G uniformly imposed horizontal seed field increased by dynamo action to $\sim 150$~G \citep{Stein2003}.

In this paper we present new realistic-type 3D radiative MHD simulations of the local solar dynamo for various seed field strengths, from $10^{-6}$ to $10^{-2}$~G (in 5 simulation runs), and investigate the development and properties of the dynamo process. In particular, we find that the magnetic field can be magnified above the equipartition strength ($\sim 600$~G), reaching more than 1000~G  in the photosphere. In the paper we discuss the initial stage of the local dynamo, formation of magnetic field strictures of different scales, location of the dynamo process, turbulent properties of magnetoconvection, as well as mechanisms and properties of spontaneously formed magnetic patches below and above the photosphere.

\section{Computational setup}
\subsection{3D radiative MHD `SolarBox' code}
We use the 3D radiative MHD code, `SolarBox', developed at NASA Ames Research Center by A. Wray and N. Mansour for performing local dynamo simulations. The code is based on a LES formulation for compressible flow, and includes a fully coupled radiation solver, in which local thermodynamic equilibrium is assumed. Radiative transfer between fluid elements is calculated using a 3D multi-spectral-bin method using long characteristics. For initial conditions we use a standard solar model of the interior structure and the lower atmosphere. The code has been carefully tested and was previously used for studying the excitation of solar acoustic oscillations by turbulent convection in the upper convection zone \citep{Jacoutot2008a,Jacoutot2008} and other problems \citep[e.g.][]{Kitiashvili2010,Kitiashvili2011,Kitiashvili2013a}.

We solve the grid-cell averaged equations for the conservation of mass (\ref{mass}), momentum (\ref{mom}), energy (\ref{energy}), and magnetic flux (\ref{eqB}):
\begin{equation} \label{mass}
\frac{\partial \rho}{ \partial t}+\left(\rho u_i\right)_{,i}=0,
\end{equation}
\begin{equation} \label{mom}
\frac{\partial \rho u_i}{ \partial t}+\left(\rho u_i u_j +
(P_{ij}+\rho\tau_{ij})\right)_{,j}=-\rho\phi_{,i},
\end{equation}
\begin{equation} \label{energy}
\frac{\partial E}{ \partial t}+
\left[E u_i + (P_{ij}+\rho\tau_{ij})u_j-(\kappa+\kappa_T) T_{,i}
+{\left(\frac{c}{4\pi}\right)}^2 \frac{1}{\sigma+\sigma_T}\left(B_{i,j}-B_{j,i} \right)B_j+
F_i^{rad}\right]_{,i}=0,
\end{equation}
\begin{equation} \label{eqB}
\frac{\partial B_i}{ \partial t}+\left[ u_j B_i - u_iB_j
-\frac{c^2}{4\pi(\sigma+\sigma_T)} \left(B_{i,j}-B_{j,i} \right)\right]_{,j}=0,
\end{equation}
where $\rho$ is the average mass density, $u_i$ is the Favre-averaged (density-weighted) velocity, $B_i$ is the magnetic field,
and $E$ is the average total energy density $E=\frac{1}{2}\rho u_i u_i + \rho e + \rho\phi +\frac{1}{8\pi}B_iB_i$, where $\phi$ is the gravitational potential and $e$ is the Favre-averaged internal energy density per unit mass. $F_i^{rad}$ is the radiative flux, which is calculated by solving the radiative transfer equation, and $P_{ij}$ is the averaged stress tensor $P_{ij}=\left(p+\frac{2}{3}\mu
u_{k,k}+\frac{1}{8\pi}B_kB_k\right)\delta_{ij}-\mu\left(u_{i,j}+u_{j,i}\right)-\frac{1}{4\pi}B_iB_j$, where $\mu$ is the viscosity. The gas pressure $p$ is a function of $e$ and $\rho$ calculated through a tabulated equation of state \citep{Rogers1996};  $\tau_{ij}$ is the Reynolds stress, $\kappa$ is the molecular thermal conductivity, $\kappa_T$ is the turbulent thermal conductivity,  $\sigma$ is the molecular electrical conductivity, and $\sigma_T$ is the turbulent electrical conductivity.

The simulation results are obtained for a computational domain of $6.4\times6.4\times6.2$~Mm, in which a 1-Mm layer of the low atmosphere is included. The grid-size is 12.5~km in the horizontal and 12~km in the vertical direction.
Above the solar surface the vertical grid size is constant and stretches out in deeper layers. The lateral boundary conditions are periodic. The top boundary is open to mass, momentum and energy fluxes, and also to the radiation flux. The bottom boundary is open only for radiation, and simulates the energy input from the interior of the Sun. Currently it is impossible to achieve a realistic Reynolds number in numerical simulations; therefore the modeling of dynamical properties of solar convection is achieved through implementation of subgrid-scale LES turbulence models. These can effectively increase the Reynolds number and provide better representation of small-scale motions. Here we used a Smagorinsky eddy-viscosity model \citep{Smagorinsky1963}, in which the compressible Reynolds stresses were calculated in the form \citep{Moin1991}: $\tau_{ij}=-2C_S\triangle^2|S|(S_{i,j}-u_{k,k}\delta_{ij}/3)+2C_C\triangle^2|S|^2\delta_{ij}/3$, where the Smagorinsky coefficients are $C_S=C_C=0.001$, $S_{ij}$ is the large-scale stress tensor, and $\triangle\equiv(dx\times dy\times dz)^{1/3}$ with $dx$, $dy$, and $dz$ being the grid-cell dimensions.

\subsection{Initial conditions}
Local dynamo action is a complicated interaction of magnetic fields and highly turbulent flows on small scales. The dynamo modeling is started by adding a very weak seed field into a hydrodynamic simulation model of fully developed solar convection. To investigate the effects of the initial seed-field properties, we consider 5 cases of magnetic field initialization (Table~\ref{tab:cases}). In three cases the initial magnetic field is $10^{-2}$~G, and has various initial distributions: ($A$) uniform vertical field, ($B$) checkerboard-like, alternating polarity patterns, and ($C$) random white noise. In case $B$ the checkerboard structure has a period of magnetic field variations of 100~km, in order to mix opposite-polarity patches in the intergranular lanes. In two other cases, $D$ and $E$, with random seed fields and strength $10^{-4}$ and $10^{-6}$~G, we test the sensitivity of the dynamo action to the initial field strength. The hydrodynamic conditions at the time of magnetic field initialization are exactly the same for the cases $B - E$.

\begin{table}[h]
 \begin{center}
 \caption{Properties of the seed magnetic field. \label{tab:cases}}
    \begin{tabular}{|c|c|c|}
        \hline
         Cases & Magnetic field & Initial field \\
          & strength, G & configuration \\
        \hline
            $A$ & $10^{-2}$ & vertical \\
            $B$ & $10^{-2}$ & checkerboard \\
            $C$ & $10^{-2}$ & white noise \\
            $D$ & $10^{-4}$ & white noise \\
            $E$ & $10^{-6}$ & white noise \\
        \hline
    \end{tabular}
 \end{center}
\end{table}

\section{Convective collapse vs the small-scale dynamo action}
\begin{figure}
\begin{center}
\includegraphics[scale=1.]{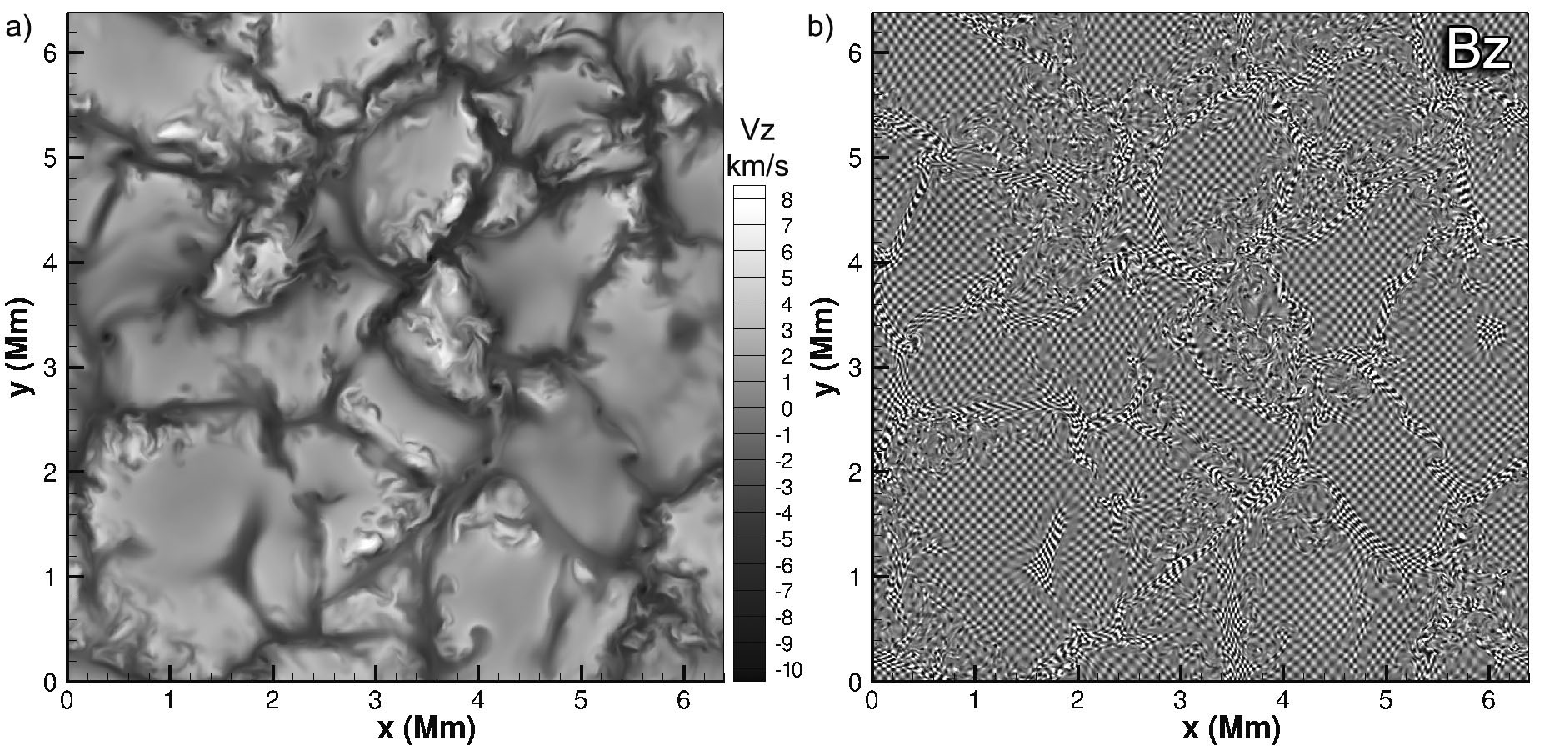}
\end{center}
\caption{Deformation of the initial checkerboard-distributed magnetic field (panel $b$) for the case $B$ (${\rm B}_{0z}=\pm 10^{-2}$~G) due to the surrounding turbulent convection (panel $a$) for $t=30$~sec after the field initialization. Panel $a$) illustrates the  distribution of the vertical velocity at the photosphere. Panel $b$) shows the deformation of the initial checkerboard structure of the
vertical magnetic field. Black-white patterns correspond to  opposite polarity magnetic fields, saturated in this image at $\pm 10^{-2}$~G. \label{fig:checker} }
\end{figure}

The initial evolution of the seed-field elements is determined by turbulent flows and demonstrates properties similar to uniformly distributed corks that tend to collect in the intergranular lanes, where the granulation flows converge. During this stage, lasting about 1 min, no dynamo action is present, and the field amplification is a result of simple compression by converging flows. Figure~\ref{fig:checker} shows an example of this effect for an initially regular checkerboard structure of magnetic field (case $B$, see Table~\ref{tab:cases}), where the deformation of the field clearly reflects a flow dynamics in which convective flow drags the magnetic field lines and helical motions cause polarity reversals on small scales.

Magnetic field amplification in turbulent solar convection can be can be roughly divided into two basic mechanisms: 1) magnetic field concentration due to converging flows, and 2) dynamo processes driven by helical or shear motions. The time delay for the natural appearance of `new' opposite-polarity patches can be estimated as a half of the overturning time on the smallest resolved scales, which for a resolution of 12~km is approximately $\sim 2$~sec. The time-lag for the appearance of a dynamo-like behavior on the smallest resolved scales due to the helical twisting of magnetic field lines can be estimated as a double overturning time on these scales, $\sim 8$~sec.

\section{Effect of small-scale dynamo action on turbulent properties of solar magnetoconvection}
Weak magnetic fields do not significantly affect dynamic properties of the quiet Sun, in terms of the life-time and size distribution of granules. Nevertheless, the dynamo-generated magnetic fields are highly inhomogeneous, and, in small-scale patches, the field strength can be high enough to act on surrounding turbulent flows (the so-called `back reaction').

In the current high-resolution simulations, which account for turbulent dynamics on sub-grid scales, we are able to capture the complicated interaction and energy exchange between the small-scale fields and flows \citep{Kitiashvili2013}.
Figure~\ref{fig:power}$a$ shows power spectra (multiplied by the \cite{Kolmogorov1941} law function $k^{5/3}$) for the turbulent kinetic energy in the photosphere for the pure hydrodynamic case (black curve), and 4 (green) and 5 (red) hours after seed-field initialization. Scaling of the spectra with $k^{5/3}$ shows the difference of the inertial range obtained from the Kolmogorov law \citep{Kolmogorov1941}.
The spectra show that, during development of the dynamo process, the growing magnetic field on small scales ($k > 10~{\rm Mm}^{-1}$) suppresses the turbulent motions that lead to kinetic energy transport to larger scales (green curve in Fig.~\ref{fig:power}$a$). The kinetic energy redistribution makes properties of turbulent flows on the solar surface closer to the theoretical Kolmogorov slope, $-5/3$. The increased magnetic flux after 5~h is strong enough to affect the turbulent flow through all scales present in the domain, which is reflected by the decreasing kinetic energy in the turbulence spectra (Fig.~\ref{fig:power}$a$, red curve).

During the first half-hour of magnetic field generation, the shape of the magnetic energy spectrum varies during an exponential growth. Later in time, the shape of the spectrum is nearly constant, showing a steady energy increase in all scales with a slope of $k^{1/3}$ for the large scales and $k^{-1}$ in the inertial range before reaching the steep dissipation range. This illustrates a second amplification phase in the dynamo process \cite[e.g.][]{Brandenburg1996} with a tendency toward an inertial-range slope of $k^{-1}$ at $t=5$~h in our simulations (Figure~\ref{fig:power}$b$).

\section{Source layers of small-scale dynamo}
It is known that magnetic field can be amplified by swirling turbulent motions, shearing and converging flows \citep[e.g.][]{Brandenburg1996,Brandenburg2012}. Magnetic field amplification due to the `stretch-twisting-fold' mechanism \citep{Childress1995} is expected to work very efficiently in regions with the strongest helical motions. Figure~\ref{fig:time} shows the vertical distribution of the rms velocity and kinetic helicity and the evolution of the mean kinetic helicity for different depths for different moments of time for the initial seed field of $10^{-6}$~G (case $E$). These distributions show that the primary layers of the local dynamo action are where the turbulent flows are strongest. The strongest helical motions occupy the top 1-Mm subsurface layer, and the helicity is transported by downdrafts from the subsurface into deeper layers. Magnetic field generated in the subsurface layer is also transported by convective downflows into deeper layers, where the field can be further compressed and amplified.

\section{Local generation of the small-scale fields}

Understanding the local small-scale magnetic field generation process is critical for studying complex solar MHD problems, such as the interaction of magnetic field and flow through different scales. To investigate field generation in detail we consider different episodes in the small-scale dynamo action.

\begin{figure}
\begin{center}
\includegraphics[scale=1.]{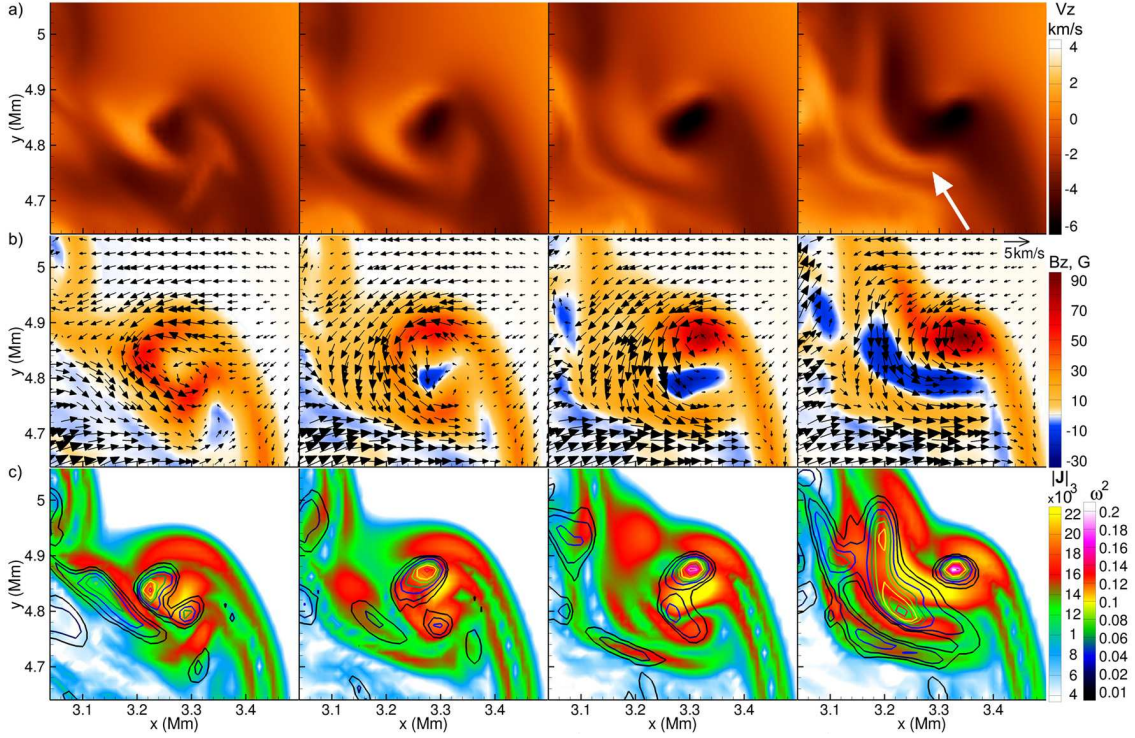}
\end{center}
\caption{Time-sequence with 15~sec cadence in a zoomed $400$~km $\times 400$~km region, where magnetic field is generated by swirling turbulent flows: $a$) vertical velocity in the photospheric layer; $b$) the vertical magnetic field evolution showing the development
of small-scale magnetic elements with opposite polarity (bipolar magnetic structure); black arrows represent the horizontal velocity field;
and c) the electric current density (background image) and the squared magnitude of vorticity ($\omega^2$, contour lines).
This example corresponds to case~$A$, with the initial $10^{-2}$~G seed field. The white arrow points to a horizontal vortex tube discussed in the text. \label{fig:case1}}
\end{figure}

The first case is demonstrated in Fig.~\ref{fig:case1}, showing the  evolution of ($a$) the vertical velocity, ($b$) the vertical magnetic field (color background) and the horizontal velocity field (arrows), and ($c$) the electric current density (color map) and enstrophy (contour lines) in a selected $400$~km $\times 400$~km region of the photosphere with 15~sec cadence.

The development of a local bipolar magnetic structure is associated with strong ($\sim 5$~km/s) swirling motions, on scales from $\sim 12-25$~km to $\sim 300-400$~km, stretching and twisting the magnetic field, which is also compressed by converging flows in the downdraft. Such such small-scale swirling motions in the intergranular lanes are typically associated with strong downflows, $\sim 6-8$~km/s  \citep{Kitiashvili2010,Kitiashvili2011}. The appearance of a bipolar magnetic structure (with a prominent negative (blue) polarity in Fig.~\ref{fig:case1}b) is also a result of swirling flows driven by a vortex tube oriented along the solar surface (indicated by the white arrow at Fig.~\ref{fig:case1}~$a$), similar to previously described by \cite{steiner2010}. The horizontally oriented vortex tube captures magnetic field lines and drags them into the subsurface layers. The dynamics of the positive polarity patch is mostly related to the vertically oriented helical motions similar to the process described by \cite{Kitiashvili2010,Kitiashvili2011}. Such complicated dynamics of flows is caused by interacting, differently oriented vortex tubes, which, in fact, have variable orientation in space, and only locally can be regarded as having `vertical' or 'horizontal' orientation \citep{Kitiashvili2012}. The generated electric current (Fig.~\ref{fig:case1}~$c$) shows a clear correlation with the swirling dynamics of convective flows. The strongest current density corresponds to areas on the periphery of the helical motions or between vortices, where stretching of the magnetic field lines is strongest.

\begin{figure}
\begin{center}
\includegraphics[scale=0.8]{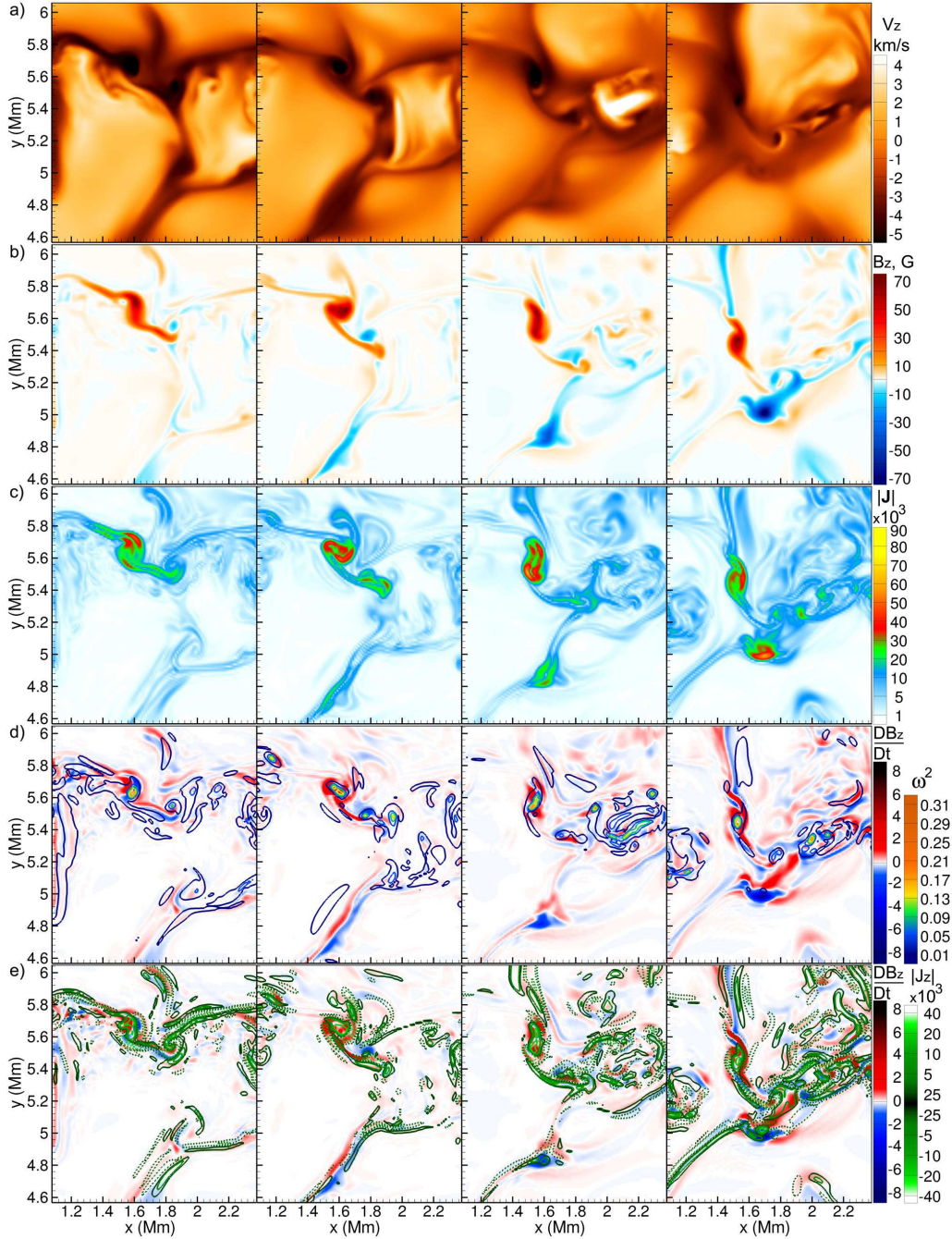}
\end{center}
\caption{Evolution of a bipolar magnetic structure in the photospheric layer illustrated in a sequence of four images with cadence 90~sec for different parameters: $a$) vertical velocity; $b$) vertical magnetic field; $c$) magnitude of the electric current density; $d$) and $e$) show the vertical magnetic field growth rate as a background red-blue image; in panel $d$) contours correspond to enstrophy and in panel $e$) contour lines show the vertical component of electric current (dashed curves for negative values). \label{fig:case3}}
\end{figure}

Another example illustrates development of a bipolar magnetic structure due to helical flows and shearing flows along the intergranular lane. Figure~\ref{fig:case3} shows a time sequence of: ($a$) the vertical velocity, ($b$) the vertical magnetic field, ($c$) the magnitude of the electric current density, ($d$) the time-derivative of the vertical magnetic field with overlaid contour lines of enstrophy, and ($e$) the vertical component of the electric current density (dashed curves correspond to negative values). An interesting feature of this example is a different scenario of the development of the magnetic elements of opposite polarity (determined by the sign of the vertical field): the positive-polarity magnetic patch evolves following the `classical' scenario of the field amplification due to helical flows described in the first example, whereas the negative-polarity patch starts forming in the intergranular lane mostly due to converging flows (magnetic collapse). The evolution of the magnetic elements of this bipolar structure is accompanied by locally growing electric current (Fig.~\ref{fig:case3}~$c, d$). Comparison of the velocity, pattern, and time derivative of the vertical magnetic field shows that some swirling motions cause magnetic field dissipation, probably due to scattering of the field lines, and often the field amplification takes place on the periphery of swirls or regions with shearing flows.

\begin{figure}
\begin{center}
\includegraphics[scale=1.]{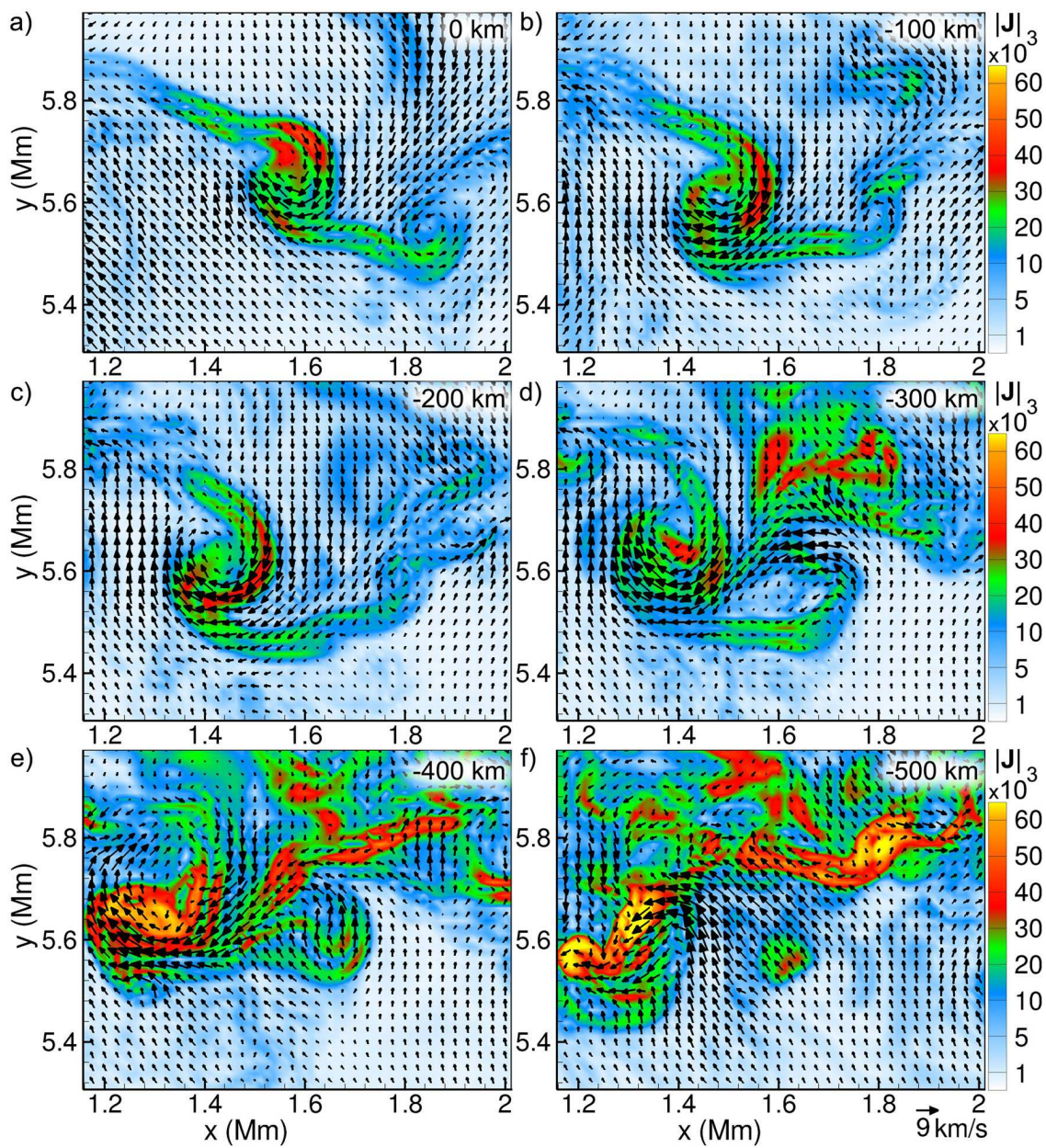}
\end{center}
\caption{The electric current density (color background) and the horizontal velocity field (arrows) at different depths,
from the photosphere (panel $a$) to 500~km below the photosphere (panel $f$) for the same moment of time as the first snapshot in Fig.~5. Arrows show the horizontal velocity field. \label{fig:case32DE-current}}
\end{figure}

\begin{figure}
\begin{center}
\includegraphics[scale=1.]{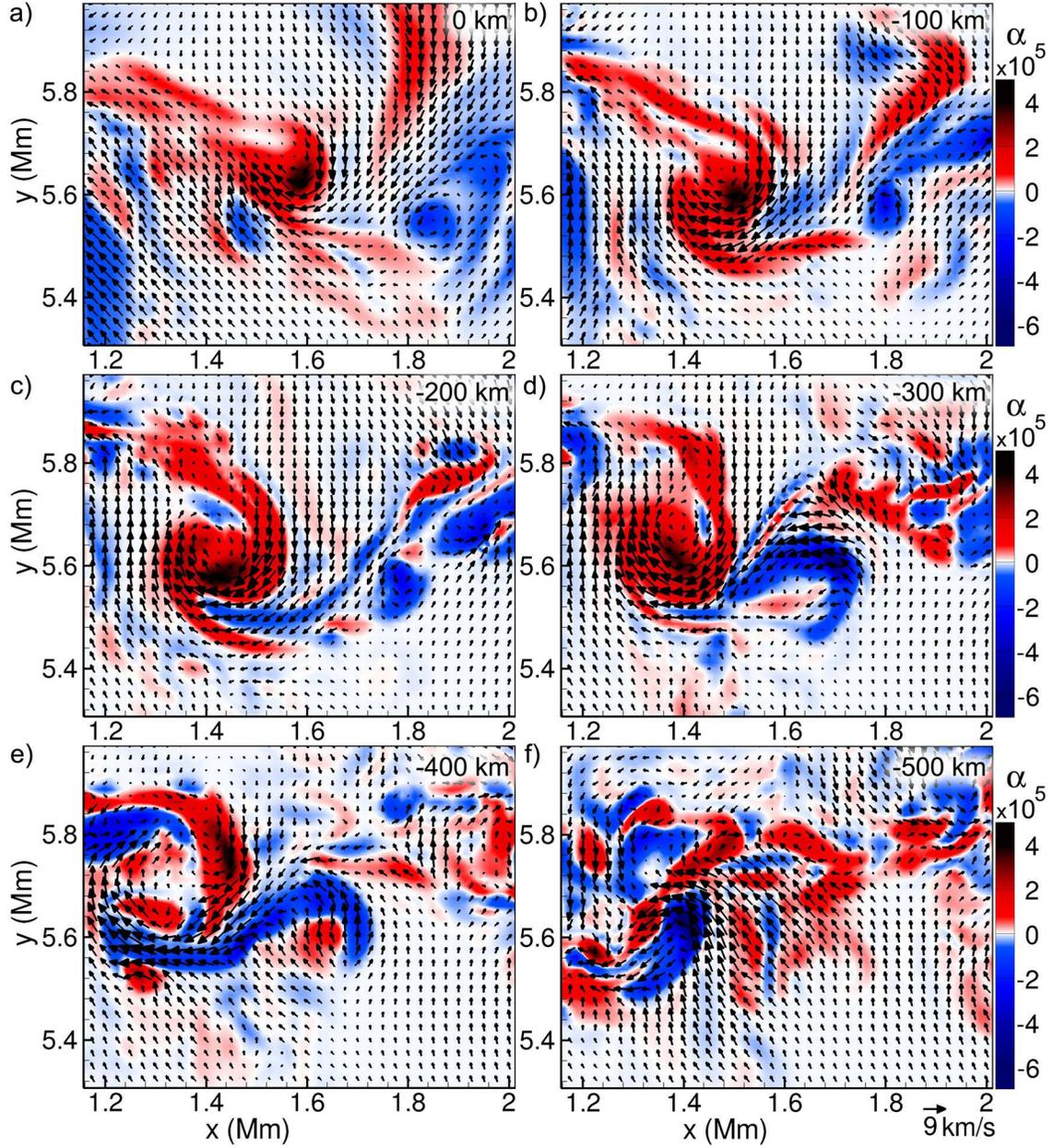}
\end{center}
\caption{The kinetic helicity, $\alpha$ (color background), and the horizontal velocity field (arrows)
at different depths from the photosphere (panel $a$) to 500~km below (panel $f$) for the same moment of time as the first snapshot in Fig.~5. \label{fig:case32DKinHel}}
\end{figure}

The distribution of vertical electric current (Fig.~\ref{fig:case32DE-current}) shows a strong correlation with the horizontal velocity of the swirling flows in the subsurface layers, and also interconnections with other subsurface current structures. In deeper layers, the magnitude of the electric current increases, but its distribution becomes diffuse and does not show a clear association with the near-surface dynamics. The kinetic helicity patterns show that the scale of the swirling motions increases with depth, from $\sim 50-60$~km at the photospheric layer (Fig.~\ref{fig:case32DKinHel}) up to $~120$~km at a depth of 300~km below the photosphere (panel $d$). In the deeper layers, the scale of helical motions continues to increase (to larger than 150~km), but the distribution of kinetic helicity becomes complicated and consists of opposite-sign helical flows, which however continue swirling together (Fig.~\ref{fig:case32DKinHel}~$e$), and disappear at a depth of about 500~km below the solar surface (panel $f$).

Cross-correlation of the kinetic helicity and the squared magnetic field strength (Fig.~\ref{fig:cross-corrHelU-B2}) shows the best correlation in subsurface layers, from the photosphere to a depth of about 500~km, indicating the primary layers where the helicity is the strongest (see also Fig.~\ref{fig:time}~$b$) and also where the generated magnetic field is transported by downflows into the deeper layers (Fig.~\ref{fig:time}~$c$).

\section{Small-scale dynamo and links to low atmosphere layers}
From previous numerical studies it is known that important dynamical and energetic links between subsurface turbulent convective flows and the low atmosphere are established through small-scale vortex tubes. In the presence of magnetic field, the vortex tubes represent channels of energy exchange between the convective layers and the chromosphere, can result in heating of chromospheric layers, and be a source of small-scale spicule-like eruptions \citep{Kitiashvili2012a,Kitiashvili2013a}. In addition to these effects, our simulations of the local dynamo show that the recently debated \citep[e.g][]{OrozcoSuarez2007,Danilovic2010,Ishikawa2010,Steiner2012,Stenflo2013} anisotropy between the vertical and horizontal magnetic field components exists only in the photosphere and above. The anisotropy changes with height but there is no dependence of these topological properties from the character of the seed field. The vertical distribution of the rms magnetic field (Fig.~\ref{fig:Brms}) shows a slowly increasing magnetic field strength with depth. In the convection zone, the distribution of the vertical and horizontal magnetic fields are similar. At the surface, the vertical magnetic field becomes dominant, and then sharply decreases above the photosphere, where the magnetic field is mostly represented by the horizontal component. This predominance of the vertical or horizontal fields in the different layers reflects the topological properties of the dynamo-generated magnetic field, which are characterized by magnetic field lines forming loops above the photosphere (Fig.~\ref{fig:B063Dm-lines}). Such topological structure resembles the magnetic canopy suggested from observations  \citep{Giovanelli1980,Jones1982,Schrijver2002}.

\begin{figure}
\begin{center}
\includegraphics[scale=0.9]{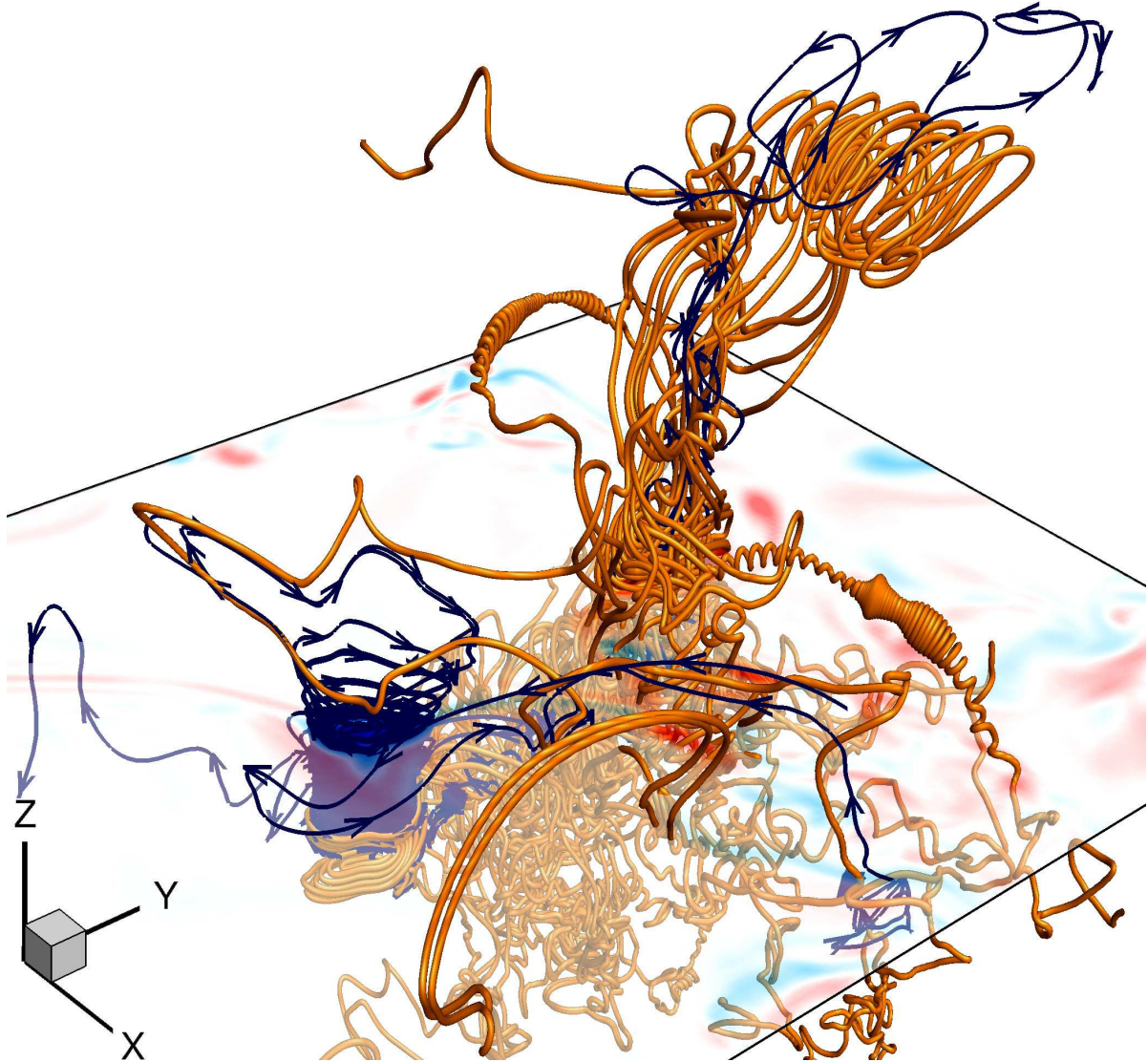}
\end{center}
\caption{Example of the topological structure of the electric current density below and above the photosphere.
Streamlines correspond to the electric currents originating from the positive (orange) and negative (blue) polarity patches. Semi-transparent horizontal plane shows the vertical magnetic field distribution in the photosphere, where blue color indicates negative polarity and red color positive polarity. The snapshot corresponds to the local dynamo event illustrated in Fig.~\ref{fig:case3}.  \label{fig:case33DE-current} }
\end{figure}

It is interesting to note that the closest opposite-polarity patches may not even be connected by magnetic field lines above the solar surface (by magnetic loops), but instead interact through electric currents above and below the photosphere. Figure~\ref{fig:case33DE-current} illustrates the topological structure by electric current density streamlines above and below the photosphere (shown as a horizontal semi-transparent plane). Each streamline is tracked from a point in the region of positive-polarity (yellow-orange streamlines) and negative polarity (dark blue) patches. The topological structure of the electric current above the photosphere is often characterized by spirals, arcs, and large swirls. Below the solar surface such topology can represent highly turbulent flows, as in the case of the positive patch (yellow streamlines, in Fig.~\ref{fig:case33DE-current}), or as a very regular spiral structure, as in the case of the negative patch (blue lines, Fig.~\ref{fig:case33DE-current}). For instance, in Figure~\ref{fig:case33DE-current} a current streamline originating in the positive polarity patch (yellow) is strongly twisted around the negative patch.

\section{Discussion and conclusion}
Observations of the Sun demonstrate the complicated dynamics of turbulent convection and its interaction with magnetic field in a wide range of temporal and spatial scales. In particular, the dynamo problem is a key to understanding variations of solar activity and various processes of self-organization. In this paper, we addressed the problem of small-scale (local) dynamos responsible for the quiet-Sun magnetic field. To investigate this problem we used a 3D radiative MHD code and performed several simulation runs for different strengths ($10^{-6}$ to $10^{-2}$~G) and spatial distributions of the initial seed field  (Table~\ref{tab:cases}).

After the initialization of the seed field, the magnetic elements first behave as corks in a fluid and start concentrating in the intergranular lanes. Shortly thereafter, overturning turbulent flows drag the field lines, and the elementary magnetic elements start changing polarity. Later, this process expands to larger scales. Because the strength of the seed field is negligible in all simulation cases, the magnetic field amplification varies among our simulation cases only during first few minutes after initialization field due to the different initial topological distributions.

\begin{figure}
\begin{center}
\includegraphics[scale=0.9]{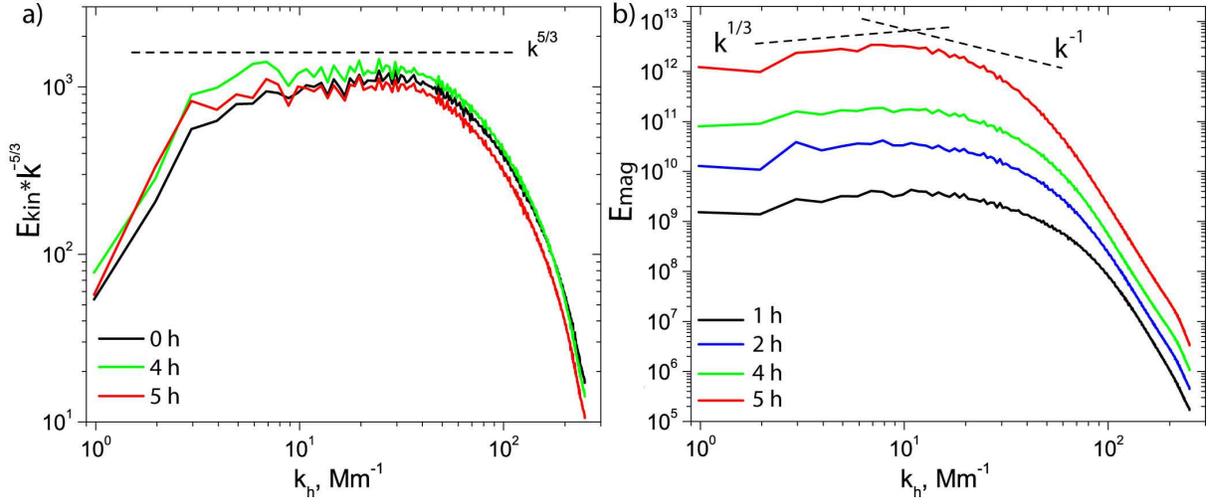}
\end{center}
\caption{Turbulent spectra for kinetic (panel $a$) and magnetic energy density (panel  $b$) in the photosphere layer (case $A$). Each spectrum is averaged over 20~min. The kinetic energy spectra are compensated by a factor $k^{5/3}$. Each curve corresponds to different moment of time: 0, 4, and 5 hours after the magnetic field initialization for the kinetic energy spectra (panel $a$); and after 1, 2, 4, and 5 hours for the magnetic energy (panel $b$). \label{fig:power}}
\end{figure}

Our simulation results show that the magnetic field amplification is driven by four primary mechanisms: 1) concentration by converging flows into the intergranular lanes, 2) shearing flows (shear dynamo), 3) helical motions (small-scale dynamo driven by the kinetic helicity, $\alpha$-effect) and 4) turbulent collapse of magnetic field. All four mechanisms of magnetic field amplification are present in our numerical model and linked to each other. Thus, the local dynamo process represents a complicated interplay of these mechanisms. We presented two characteristic examples: one in which a bipolar magnetic structure was generated by the interaction of vertical and horizontal vortex tubes, and another in which the field was generated by a combination of a vertical vortex tube and shearing flows. The flow topology of the local dynamo patches is very complicated and requires further investigation. However, the primary topology of the dynamo-generated magnetic field is represented by compact magnetic loops appearing as bipolar structures in the intergranular lanes and forming an `internetwork'. Development of such a magnetic network causes redistribution of the turbulent kinetic energy by transferring kinetic energy from small to large scales, bringing the turbulent spectrum of the photosphere closer to the Kolmogorov power law, $k=-5/3$ (Fig.~\ref{fig:power}).

\begin{figure}
\begin{center}
\includegraphics[scale=1.]{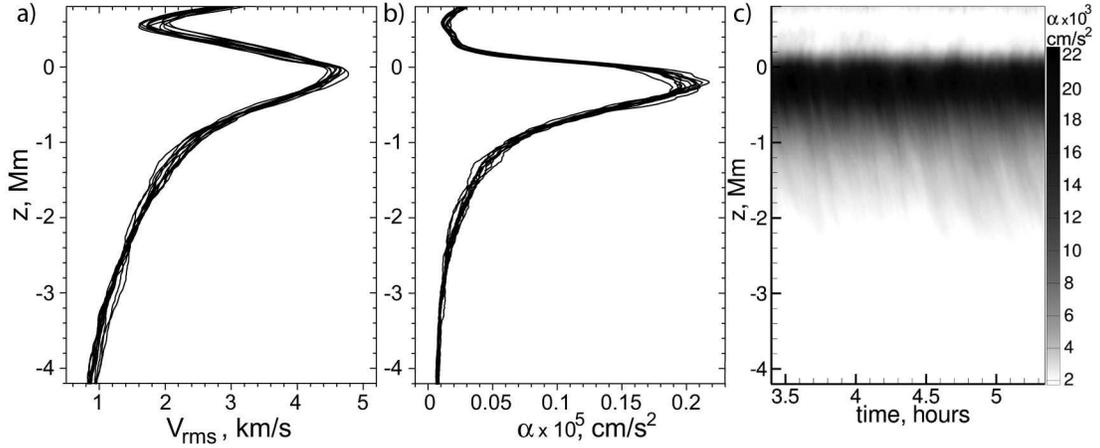}
\end{center}
\caption{Vertical profiles of rms velocity (panel $a$) and kinetic helicity ($b$) with  5~min cadence  5~h after the field initialization. Panel $c$ shows the time evolution of the mean helicity at different depths for case $E$ (Table 1) with the initial seed field of $10^{-6}$~G. \label{fig:time}}
\end{figure}

\begin{figure}
\begin{center}
\includegraphics[scale=1]{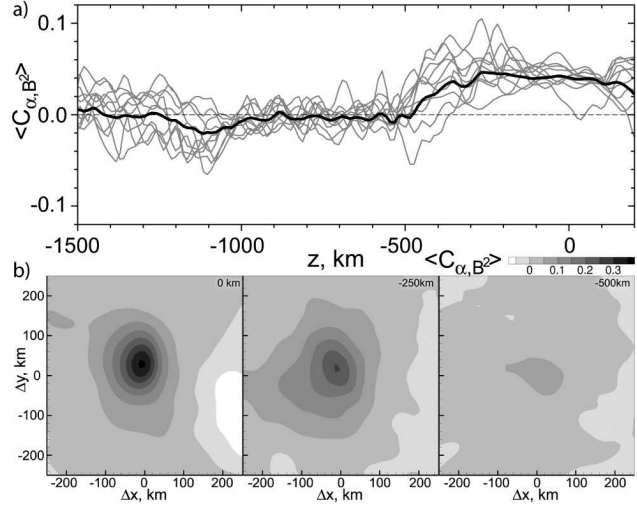}
\end{center}
\caption{The mean profiles of cross-correlation between the kinetic helicity, $\alpha$, and squared magnetic field, $B^2$, as a function of depth. Gray thin curves correspond to different moments of time for the case illustrated in Fig.~5, with 30~sec cadence. Black curve is the time-averaged cross-correlation function. \label{fig:cross-corrHelU-B2}}
\end{figure}

The process of amplification of magnetic field occupies the upper 500~km-deep subsurface layer, where the helical flows are strongest (Fig.~\ref{fig:time}$b$,~\ref{fig:cross-corrHelU-B2}). The dynamo-generated magnetic field is then transported by convective downflows into deeper layers (Fig.~\ref{fig:time}~$c$).

\begin{figure}
\begin{center}
\includegraphics[scale=1]{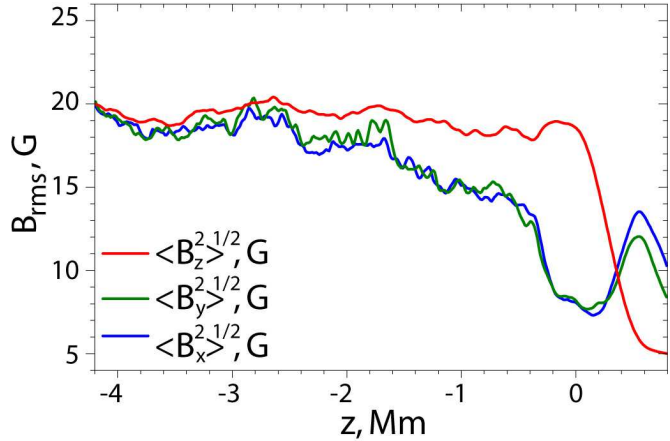}
\end{center}
\caption{One-hour time-averaged rms magnetic field profiles: $<B_{x}^2>^{1/2}$ (blue curve), $<B_{y}^2>^{1/2}$ (green) and $<B_{z}^2>^{1/2}$ (red curve) as a function of depth. \label{fig:Brms}}
\end{figure}

\begin{figure}
\begin{center}
\includegraphics[scale=1.]{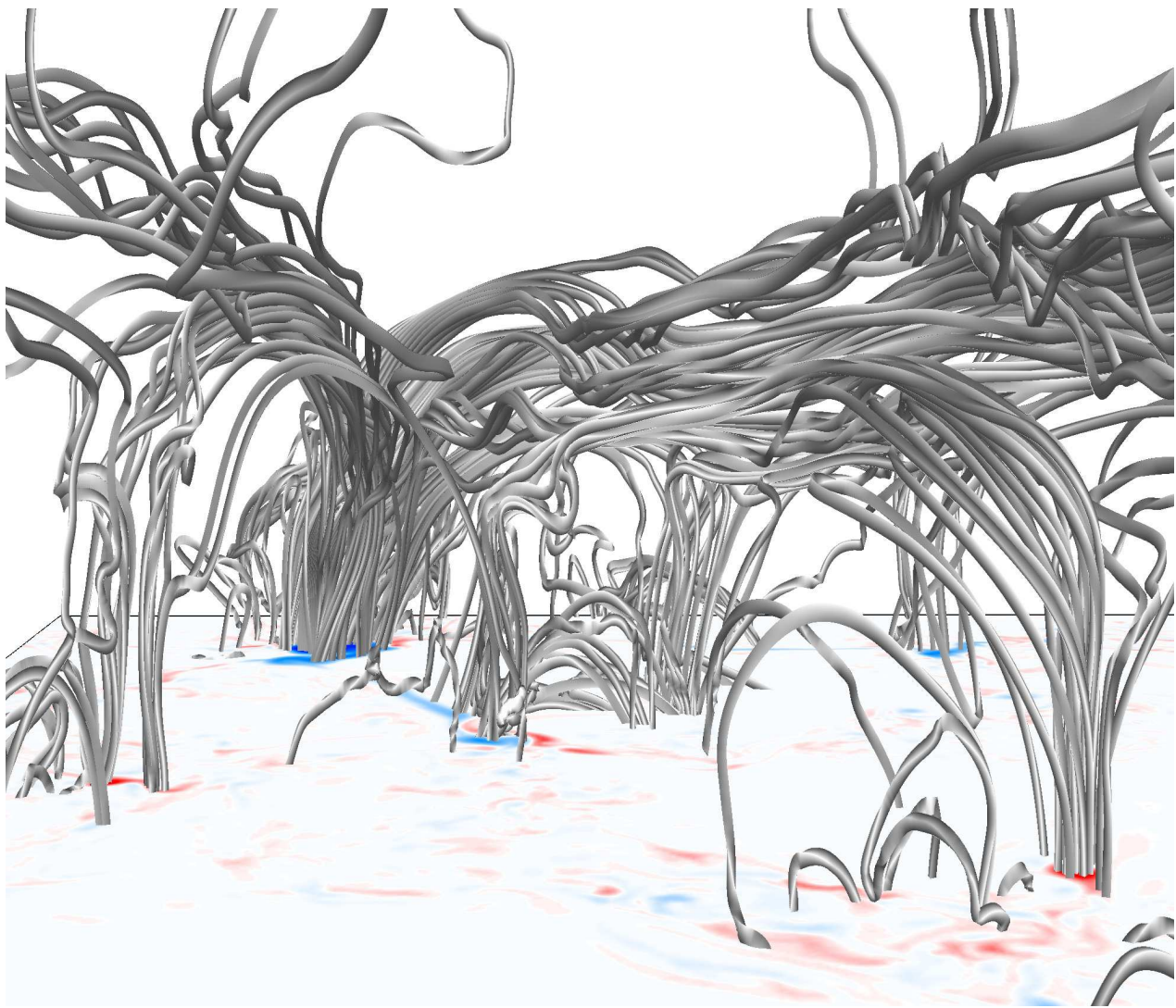}
\end{center}
\caption{Topology of the magnetic field lines above the photosphere in the local dynamo simulations. The horizontal plane shows the distribution of the vertical magnetic field in the photosphere. Red color corresponds to positive polarity, blue color to negative polarity of the vertical magnetic field. The range of field strength is from $-800$~G to 300~G. \label{fig:B063Dm-lines}}
\end{figure}

Our simulations show that the relative ratio between the vertical and horizontal field components (field anisotropy) changes
with height (Fig.~\ref{fig:Brms}). The vertical fields are dominant near the photosphere, but the horizontal fields become
much stronger than the vertical fields in the atmosphere. This can be explained by the loop-like topology of magnetic field
(Fig.~\ref{fig:B063Dm-lines}). Perhaps, this change of the anisotropy of the vertical and horizontal fields can explain
discrepancies among different observations \citep[e.g][]{OrozcoSuarez2007,Danilovic2010,Ishikawa2010,Stenflo2013} and resolve the controversy.

\end{document}